\documentclass[aps,prl,preprint,showpacs,showkeys,groupedaddress]{revtex4}
\usepackage{graphicx}
\usepackage{amssymb}
\usepackage{amsmath}
\usepackage{amsfonts}
\begin{document}
\title{Transport properties of Brownian particles confined to a narrow channel by a periodic potential}
\author{Xinli Wang}
\author{German Drazer}
\affiliation{Department of Chemical and Biomolecular Engineering \\ The Johns Hopkins University, Baltimore, MD 21218 USA}
\date{\today}
\begin{abstract}
We investigate the transport of Brownian particles in a two-dimensional potential under the action of a uniform external force.
The potential is periodic in one direction and confines the particle to a narrow channel of varying cross-section
in the other direction.
We apply the standard long-wave asymptotic analysis in the narrow
dimension and show that the leading order term is equivalent to that obtained previously from a direct extension
of the Fick-Jacobs approximation. We also show that the confining potential has similar effects on the transport of Brownian particles
to those induced by a solid channel. Finally, we compare the analytical results with Brownian dynamics simulations in the case
of a sinusoidal variation of the width of the parabolic potential in the cross-section. We obtain excellent agreement for the marginal
probability distribution, the average velocity of the Brownian particles and the asymptotic dispersion coefficient,
over a wide range of P\'eclet numbers.
\end{abstract}
\maketitle

\section{Introduction}
Recent progress in microfluidic devices has led to the development of novel separation strategies that
take advantage of the unprecedented control on the geometry and chemistry of the stationary phase at
scales that are comparable to the size of the transported species \cite{Duke98,EijkelB06,Pamme07}.
A fundamental problem that is at the core of several of the proposed separation devices is the transport of
Brownian particles through entropy barriers. A representative example is the motion of a suspended
particle in a channel with periodically varying cross section.
The purely-diffusive transport in the absence of an
external force has been studied extensively \cite{Zwanzig92,RegueraR01,KalinayP05,BuradaHMST09}, and a
well-known approach is to reduce the dimensionality of the problem via the Fick-Jacobs approximation.
In this approximation the motion in the cross section is reduced to an entropic contribution to the longitudinal transport \cite{Jacobs67}.
The case of biased diffusion in the presence of a driving force has also
received considerable attention due to its relevance to separation devices.
H\"anggi and coworkers examined the validity of a direct extension of the Fick-Jacobs equation to describe the
biased motion of a Brownian particle in a narrow channel of varying cross section. The authors showed that
using an ad-hoc position-dependent diffusivity provides a good approximation, for P\'eclet numbers below a critical
value \cite{RegueraSBRRH06,BuradaSRRH07}. Dorfman and coworkers \cite{LaachiKYD07} also investigated the
biased motion of a Brownian particle in a periodic channel. Specifically, they performed an asymptotic perturbation
analysis to obtain simple expressions for the macroscopic transport coefficients that remain valid at relatively large P\'eclet numbers.

Here, we also consider the transport of Brownian particles confined to a channel of periodically varying cross section but,
in the present case, the confinement is induced by a potential energy landscape and not the solid boundaries of a channel.
This type of spatial confinement occurs, for example, in the transport of suspended particles in microfluidic channels.
Depending on the density of the particles, the van der Waals and electrostatic forces between the particle and the channel walls,
and the chemical composition of the media, the particles could become confined to the secondary minima of the particle-wall interaction
potential in the vertical direction \cite{RusselSS89}.
In that case, the presence of a periodic pattern on the bottom wall, such as that created by the deposition of thin metal stripes perpendicular
to the flow, would lead to the confinement of the Brownian particles to a periodically varying channel parallel to the wall,
analogous to the one considered here.
We shall show that the effect on the average velocity of the particles (and the probability distribution in general) induced by this type of {\it soft} confinement
of the suspended particles to an energy minima is the same as that caused by geometric confinement between solid walls.
Understanding the effect that this type of confinement has on the transport of suspended particles is particularly important
for the development of recently proposed separation techniques in microfluidic devices that are based on partitioning \cite{DorfmanB01}.

\section{Transport of Brownian particles in a confining periodic potential}

Let us consider the transport of Brownian particles in a potential that is periodic in the $x$-direction,
$V(x=x_0,z)=V(x=x_0+L,z)$, and that confines the particles in the $z$-direction, that is $V(x,z)\to+\infty$ for $z\to \pm \infty$.
In the limit of negligible inertia effects the motion of the particles is described by the Smoluchowski equation for the probability density $P(x,z,t)$,
\begin{equation}
\frac{\partial P}{\partial t}+\nabla\cdot {\bf J}=\delta(x,z)\delta(t).
\end{equation}
The probability flux, ${\bf J}(x,z,t)$, is given by
\begin{equation}
{\bf J} = \frac{1}{\eta}\left(FP - \frac{\partial V}{\partial x}P - k_{B}T\frac{\partial P}{\partial x}\right)\vec{i} + \frac{1}{\eta}\left(-\frac{\partial V}{\partial z}P-k_{B}T\frac{\partial P}{\partial z}\right)\vec{k},
\end{equation}
where $F$ is a uniform external force in the $x$-direction, $\eta$ is the viscous friction coefficient,
and we have used the Stokes-Einstein equation to write the diffusion coefficient as a function of $\eta$, $D=k_{B}T/\eta$.

In order to obtain the asymptotic distribution of particles within a single period of the potential we first introduce
the reduced probability density and the reduced probability current
(see Refs. \cite{Reimann02, LiD07} or the analogous approach presented in Ref. \cite{BrennerE93}),
\begin{eqnarray}
\label{reduced}
\tilde P(x,z,t)= \sum_{n_x=-\infty}^{+\infty} P(x+n_x L,z,t), \\
\mathbf{\tilde J}(x,t)= \sum_{n_x=-\infty}^{+\infty}\mathbf{J}(x+n_xL,z,t).
\end{eqnarray}
The reduced probability is obtained by solving the Smoluchowski equation with periodic boundary conditions in $x$.
In particular, the long-time asymptotic probability density, $P_{\infty}(x,z)=\lim _{t\to\infty}\tilde P(x,z,t)$, is governed by the equation
(dropping the tilde),
\begin{equation}
\nabla\cdot {\bf J}_{\infty} = 0.
\label{eqn:governing_original}
\end{equation}
The far-field condition in $z$ is a vanishingly small probability density and flux due to the confining potential,
\begin{equation}
J_{\infty}^{z} = \frac{1}{\eta}\left(-\frac{\partial V}{\partial z}P_{\infty}-k_{B}T\frac{\partial P_{\infty}}{\partial z}\right) \xrightarrow[{z \to \pm \infty}]{}  0.
\label{eqn:flux}
\end{equation}
Finally, the reduced probability is obtained by imposing the periodic boundary conditions in $x$,
\begin{equation}
P_{\infty}(x=0,z)=P_{\infty}(x=L,z),
\label{eqn:period}
\end{equation}
 and the normalization condition,
\begin{equation}
\left<P_{\infty}\right>=\int_{0}^{L}dx\int_{-\infty}^{\infty}P_{\infty}dz = 1.
\label{eqn:normalization}
\end{equation}

In the case of narrow or {\it slender} channels, we assume that the characteristic length scale
in the cross-section, i. e. perpendicular to the channel centerline, is given by $\epsilon L$, where $\epsilon \ll 1$ is the
slenderness ratio. We can then write the governing equation and boundary conditions using the following dimensionless variables,
$\overline{x}=x/L$, $\overline{z} = z/(\epsilon L)$, $\overline{V} = V/(k_B T)$, where $k_B$ is the Boltzmann constant and
$T$ is the absolute temperature, and the re-scaled probability density $\overline{P}_{\infty}=\epsilon L^{2}P_{\infty}$.
For the sake of simplicity all bars are dropped after nondimensionalization. The governing equation for the reduced probability
(Eq. (\ref{eqn:governing_original})) then becomes:
\begin{equation}
\epsilon^{2}\frac{\partial}{\partial x}\left[\left(\textrm{Pe}-\frac{\partial V}{\partial x}\right)P_{\infty}-\frac{\partial P_{\infty}}{\partial x}\right] + \frac{\partial}{\partial z}\left[-\frac{\partial V}{\partial z}P_{\infty}-\frac{\partial P_{\infty}}{\partial z}\right] = 0,
\label{eqn:governing}
\end{equation}
where the P\'eclet number is defined as $\textrm{Pe}=FL/k_B T$.
Simple inspection of this equation shows that the leading order approximation for $\epsilon \ll 1$
is the local equilibrium in $z$, which corresponds to $J_{0}^{z}(x,z)=0$, due to the no-flux far-field condition.
In general, an accurate description of the long-time transport of Brownian particles can be obtained from the first two moments of the
asymptotic probability, which describe the average velocity and the broadening of the distribution.
Applying macrotransport theory we can write the average velocity in terms of the asymptotic probability distribution \cite{BrennerE93},
\begin{equation}
\langle v\rangle=\int_{0}^{1}dx\int_{-\infty}^{\infty}J_\infty^{x}dz=\int_{0}^{1}dx\int_{-\infty}^{\infty}
\left[\left(\textrm{Pe}-\frac{\partial V}{\partial x}\right)P_{\infty}-\frac{\partial P_{\infty}}{\partial x}\right]dz.
\end{equation}
The dispersion coefficient $D^{*}$, can also be calculated from the asymptotic distribution via the so-called $B-$field,
which is the solution of the following differential equation \cite{BrennerE93},
\begin{equation}
\label{Eq:B}
\begin{split}
&\frac{\partial}{\partial z}\left(P_{\infty}\frac{\partial B}{\partial z}\right)-\left(-\frac{\partial V}{\partial z}P_{\infty}-\frac{\partial P_{\infty}}{\partial z}\right)\frac{\partial B}{\partial z}\\&\quad+\epsilon^{2}\left[\frac{\partial}{\partial x}\left(P_{\infty}\frac{\partial B}{\partial x}\right)-\left(\left(\textrm{Pe}-\frac{\partial V}{\partial x}\right)P_{\infty}-\frac{\partial P_{\infty}}{\partial x} \right)\frac{\partial B}{\partial x}\right]=\epsilon^{2}P_{\infty}\langle v\rangle.
\end{split}
\end{equation}
The boundary conditions for the $B-$field are,
\begin{eqnarray}
\frac{\partial B}{\partial z} \xrightarrow[{z \to \pm \infty}]{}0, \\ \nonumber
B(x=1,z)-B(x=0,z)=-1.
\end{eqnarray}
Finally, the dispersion coefficient is given in terms of $B(x,z)$ by
\begin{equation}
\label{D*}
D^{*}=\int_{0}^{1}dx\int_{-\infty}^{\infty}P_{\infty}\left[\left(
\frac{\partial B}{\partial x}\right)^{2}+\frac{1}{\epsilon^{2}}\left(\frac{\partial B}{\partial z}\right)^{2}\right]dz.
\end{equation}

\section{Asymptotic analysis in the Narrow channel approximation}
We apply the standard long-wave asymptotic analysis to obtain an approximate solution to the problem described above.
First, we propose a solution to the stationary probability distribution in the form of a regular perturbation expansion
in the slenderness parameter,
\begin{equation}
P_{\infty}(x,z) \sim p_{0}+\epsilon^{2}p_{1}+\epsilon^{4}p_{2}+\cdots.
\end{equation}
Analogously, we write a regular perturbation expansion for the probability flux,
\begin{equation}
{\bf J}_{\infty}(x,z) \sim {\bf J}_{0}+\epsilon^{2}{\bf J}_{1}+\epsilon^{4}{\bf J}_{2}+\cdots.
\end{equation}

Substituting these expansions into Eq. (\ref{eqn:governing}) it is straightforward to determine the governing equation for the
leading order terms,
\begin{equation}
\frac{\partial}{\partial z}\left(-\frac{\partial V}{\partial z}p_{0}-\frac{\partial p_{0}}{\partial z}\right)=\frac{\partial J_{0}^{z}}{\partial z}=0,
\label{eqn:governing_0}
\end{equation}

The corresponding leading order boundary and normalization conditions, derived from Eqs. (\ref{eqn:flux}-\ref{eqn:normalization}), are:
\begin{eqnarray}
J_{0}^{z}\left(x,z=\pm\infty\right) &=& 0 \\ \nonumber
p_{0}\left(x=0,z\right) &=& p_{0}\left(x=1,z\right) \\ \nonumber
\left<p_{0}\right> &=& 1.
\end{eqnarray}
Then, integrating Eq. (\ref{eqn:governing_0}) and taking into account the no-flux condition we obtain:
\begin{equation}
p_{0}(x,z)=f_0(x)e^{-V(x,z)},
\label{eqn:p0}
\end{equation}
where $f_0(x)$ is an unknown function that is to be determined from the second order $O(\epsilon^{2})$ balance of
Eq. (\ref{eqn:governing}),
\begin{equation}
\frac{\partial}{\partial z}\left(\frac{\partial V}{\partial z}p_{1}+\frac{\partial p_{1}}{\partial z}\right)=\frac{\partial}{\partial x}\left[\left(\textrm{Pe}-\frac{\partial V}{\partial x}\right)p_{0}-\frac{\partial p_{0}}{\partial x}\right].
\label{eqn:governing_1}
\end{equation}
The corresponding boundary and normalization conditions are:
\begin{eqnarray}
J_{1}^{z}\left(x,z=\pm\infty\right) &=& 0, \\ \nonumber
p_{1}\left(x=0,z\right) &=& p_{1}\left(x=1,z\right) \\ \nonumber
\left<p_{1}\right> &=& 0.
\end{eqnarray}
Substituting the solution obtained for $p_{0}$ into Eq. (\ref{eqn:governing_1}) we obtain
\begin{equation}
\frac{\partial}{\partial z}\left(\frac{\partial V}{\partial z}p_{1}+\frac{\partial p_{1}}{\partial z}\right)=\frac{\partial}{\partial x}\left[\left(\textrm{Pe}f_0-\frac{df_0}{dx}\right)e^{-V(x,z)}\right].
\label{eqn:p1B0}
\end{equation}
In steady state, the total flux in the $x$-direction is constant along the channel. Therefore,
by integrating the equation above over the cross-section, and taking into account the zero flux condition
in the $z$-direction, we obtain
\begin{equation}
\frac{d}{dx}\left[\left(\textrm{Pe}f_0-\frac{df_0}{dx}\right) I(x)\right]=0,
\label{eqn:B0}
\end{equation}
where
\begin{equation}
I(x)=\int_{-\infty}^{\infty}e^{-V(x,z)}dz.
\end{equation}

We note that $I(x)$ plays the role of the width $w(x)$ of a solid channel, that is, the integral in the previous equation becomes equal to the width of
the channel if we model the rigid boundaries by a potential field that is zero (infinite) inside (outside) the channel.
In fact, replacing $I(x)$ by $w(x)$ in Eq. (\ref{eqn:B0}) we obtain Eq. (25) in Ref. \cite{LaachiKYD07}. This suggests that the confining
potential $V(x,z)$ has similar effects on the transport of Brownian particles to those induced by a channel with solid walls.

Before we proceed to solve the equation for $f_0(x)$ it is also interesting to compare it with the equation obtained from the direct extension of
the Fick-Jacobs approximation in the presence of an external force. In fact, in the narrow geometry that we are considering here, we can assume that the particle will reach local equilibrium in the cross section fairly rapid compared to its diffusive or convective motion along the channel.
This separation of time scales suggests that the conditional probability $P(z/x,t)$ could be approximated by the equilibrium
distribution $P_{eq}(z/x)$. Then, the total probability density takes the form,
\begin{equation}
P(x,z,t)=P(z/x,t) P(x,t)\approx P_{eq}(z/x) p(x,t)= \frac{e^{-V(x,z)}}{I(x)}p(x,t),
\end{equation}
where $p\left(x,t\right)$ is the marginal probability distribution,
\begin{equation}
p(x,t)=\int_{-\infty}^{\infty}P(x,z,t)dz,
\end{equation}
Considering the long time limit, we can then use the same approximation for the asymptotic distribution,
\begin{equation}
P_{\infty}(x,z)\approx  \frac{e^{-V(x,z)}}{I(x)}p(x).
\end{equation}
Finally, we can project the problem into the longitudinal direction by substituting this expression into the governing equation
for the probability density and integrating over the cross-section,
\begin{equation}
\frac{d}{dx}\left\{\left[\textrm{Pe}\left(\frac{p}{I(x)}\right)-\frac{d}{d x}\left(\frac{p}{I(x)}\right)\right]I(x)\right\}=0.
\label{eqn:FJ}
\end{equation}

This is the form of the Fick-Jacobs equation used in the presence of an external force in previous studies
\cite{RegueraR01, RegueraSBRRH06}. It is interesting to point out that, by making the substitution $p(x)=f_0(x)I(x)$,
the previous equation becomes identical to Eq. (\ref{eqn:B0}),
which was derived from the asymptotic analysis. Therefore, the leading order term of the asymptotic analysis is equivalent to the
the proposed extension of the Fick-Jacobs approximation in the presence of an external field.
Then, the validity of the ad-hoc position-dependent effective diffusivity proposed in Ref. \cite{BuradaSRRH07} could, in principle,
be tested by extending the present approach to higher orders in the slenderness parameter.

We now go back to equation (\ref{eqn:B0}) to obtain the general solution for $f_0(x)$,
\begin{equation}
f_{0}(x)=e^{\textrm{Pe}\,x}\left(C_{1}\int\frac{e^{-\textrm{Pe}\,x}}{I(x)}dx+C_{2}\right),
\end{equation}
where $C_{1}$ and $C_{2}$ are the constants of integration. The leading order contributions to the
average velocity and the dispersion coefficient can then be calculated from the probability distribution.
First, we obtain the general expression for the average velocity following a derivation analogous to that
presented in Ref. \cite{BuradaSRRH07} (appendix A) for the particle current,
\begin{equation}
\left<v\right>=(1-e^{-\textrm{Pe}})\left[\int_{0}^{1}dxe^{\textrm{Pe}\,x}I(x)\int_{x}^{x+1}
\frac{dx^{'}}{e^{\textrm{Pe}\,x^{'}}I(x^{'})}\right]^{-1}.
\end{equation}

In order to calculate the dispersion coefficient given in Eq. (\ref{D*}) we first propose a
regular perturbation expansion for the $B$-field, analogous to those proposed for the probability
density and flux,
\begin{equation}
B\sim B_{0}+\epsilon^{2}B_{1}+\epsilon^{4}B_{2}+\cdots.
\end{equation}
Substituting the expansion for both the probability density and the $B$-field into Eq. (\ref{Eq:B})
it is straightforward to show that the zeroth-order balance is an homogeneous equation that is
satisfied by an arbitrary function $B_0(x)$. The governing equation for $B_0(x)$ is obtained
from the $O(\epsilon^{2})$ balance, after integrating over the cross-section and taking into
account the no-flux boundary condition for the $B$-field in the $z$-direction,
\begin{equation}
\frac{d^2B_0}{dx^2}+\left(\frac{p_0'-\langle v\rangle}{p_0}\right)\frac{dB_0}{dx}=\langle v\rangle.
\label{eqn:Bfield}
\end{equation}
The remaining boundary condition is,
\begin{equation}
B_0(1)-B_0(0)=-1.
\label{eqn:Bfield_BC}
\end{equation}
The above two equations determine $B_0$ uniquely to within an arbitrary additive constant,
which does not affect the calculation of the effective dispersion coefficient \cite{BrennerE93},
\begin{equation}
D^*=\int_0^1 dx\int_{-\infty} ^{\infty} p_0(x,z)\left(\frac{dB_0}{dx}\right)^2dz.
\end{equation}

\section{Transport of Brownian particles confined by a parabolic potential}
\begin{figure}[!ht]
\centering
\includegraphics[width=4in]{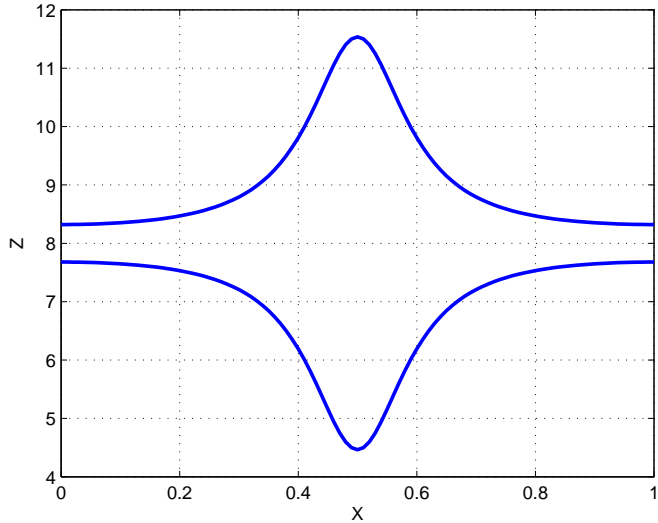}
\caption{Equipotential lines corresponding to $z=z_0=\pm 2\delta(x)$ for a periodic potential
$V(x,z)=(\cos 2\pi x + d)^{2}(z-z_{0})^{2}$ with $d=1.2$ and $z_{0}=8$.}
\label{fig:channel}
\end{figure}

In this section, we discuss a specific example of the confining potential that illustrates our previous results and allows direct comparison
with numerical results. Consider the transport of Brownian particles confined to a narrow channel by a parabolic potential of periodically
varying width,
\begin{equation}
 V(x,z)= \left(\cos \frac{2\pi}{L} x+d\right)^{2}\left(\frac{z-z_{0}}{\epsilon L}\right)^{2} k_B T,
\label{eqn:potential_0}
\end{equation}
where $z_0$ is the position of the center of the channel, which we assume to be constant (straight centerline), and $d>1$ determines
the minimum opening of the channel. The slenderness of the geometry is given by $\epsilon\ll 1$, which is the ratio of the characteristic
width of the potential in the $z$-direction, $\epsilon L$, to the length of one period, $L$.
Using the same dimensionless variables introduced before and dropping again all bars for the sake of simplicity, we obtain,
\begin{equation}
V(x,z) = \frac{\left(z-z_{0}\right)^{2}}{2\delta^{2}\left(x\right)},
\label{eqn:potential}
\end{equation}
where
\begin{equation}
\delta\left(x\right) = \frac{1}{\sqrt{2}\left(\cos 2\pi x +d\right)}.
\end{equation}
It is clear then that the particles will be confined in the $z$ direction by a parabolic potential and that the
width of the confining region is determined by $\delta(x)$.
In fact, the equilibrium distribution of particles is given by the Boltzmann distribution with
variance $\sigma(x)=\delta(x)$. In Fig. \ref{fig:channel} we plot the equipotential
lines $z-z_0 = \pm 2\delta(x)$, with $d = 1.2$ and $z_{0}=8$. In equilibrium, approximately $95\%$ of the particles
are confined to the region enclosed by these lines.

For this specific potential we obtain $I(x)=\sqrt{2\pi}\delta(x)=\sqrt{\pi}(\cos2\pi x+d)^{-1}$, and
\begin{equation}
f_0(x)=-\frac{C_{1}}{\sqrt{\pi}}\left(\frac{2\pi\sin2\pi x-\textrm{Pe}\,\cos2\pi x}{4\pi^{2}+(\textrm{Pe})^{2}}-\frac{d}{\textrm{Pe}}\right)+C_{2}\, e^{\textrm{Pe}\,x},
\end{equation}
Where $C_1$ and $C_2$ are integration constants. In particular, $C_1$ is the first constant of integration of  Eq. (\ref{eqn:B0}),
which corresponds to the total flux in the $x$-direction, $C_1=\langle J^x_0 \rangle$.
Thus, imposing periodicity and the normalization condition we obtain the leading order probability density,
\begin{equation}
p_{0}(x,z)= -\frac{\langle J^x_0 \rangle}{\sqrt{\pi}}e^{-V(x,z)}\left(\frac{2\pi\sin2\pi x-\textrm{Pe}\cos2\pi x}{4\pi^{2}+\textrm{Pe}^{2}}-\frac{d}{\textrm{Pe}}\right),
\label{eqn:p0_analytical}
\end{equation}
and average velocity,
\begin{equation}
\label{avV}
\langle J^x_0 \rangle = \langle v \rangle = \frac{\left(\textrm{Pe}^{2}+4\pi^{2}\right)\textrm{Pe}}{\textrm{Pe}^{2}+\frac{4\pi^{2}d}
{\sqrt{d^{2}-1}}}.
\end{equation}
Finally, the leading order term of the marginal probability distribution, corresponding to the distribution in the heuristic extension to the
Fick-Jacobs approximation, is given by
\begin{equation}
\label{marginal}
p(x)=\frac{\textrm{Pe}^2}{\textrm{Pe}^{2}+\frac{4\pi^{2}d}{\sqrt{d^{2}-1}}}\left[ 1 - \frac{1}{\textrm{Pe}^{2}}
\frac{2 \pi \textrm{Pe} \sin2\pi x-4\pi^2d}{d+\cos(2\pi x)}\right]
\end{equation}
Note that in the limiting case in which diffusive transport is dominant (vanishingly small P\'eclet number) we recover the equilibrium Boltzmann distribution,
\begin{equation}
\lim_{\mbox{Pe}\to 0}p_{0}(x,z)=\sqrt{\frac{d^{2}-1}{\pi}}e^{-V(x,z)}.
\end{equation}
On the other hand, in the limit of deterministic motion we obtain,
\begin{equation}
\lim_{\mbox{Pe}\to \infty}p_{0}(x,z)=\frac{d+\cos(2\pi x)}{\sqrt{\pi}}e^{-V(x,z)}.
\end{equation}
In this case, there is no boundary layer developing due to the large magnitude of the driving force, as it would be the case in the presence of solid
walls and permeating forces \cite{LiD07}. Therefore, the diffusive fluxes are negligible compared to convective transport,
and the average velocity tends to its bulk value, something that is also clear from Eq. (\ref{avV}).
In addition, the integral of the asymptotic distribution over a cross section is uniform, which is also clear from Eq. (\ref{eqn:B0}).
Note, however, that this limit is valid if the slenderness condition satisfies $\epsilon^2 \textrm{Pe} \ll 1$, as discussed in Ref. \cite{LaachiKYD07}.


\section{Brownian dynamics simulations}
We performed Brownian dynamics simulations of particle transport under confinement by a potential landscape
to examine the previous asymptotic results in more detail.
The motion of Brownian particles in a viscous solvent in the limit of vanishingly small inertia
is governed by the overdamped Langevin equations,
\begin{equation}
\eta\frac{dx}{dt}=F-\frac{\partial V}{\partial x}+\sqrt{\eta k_{B}T}\zeta(t),
\end{equation}
and
\begin{equation}
\eta\frac{dz}{dt}=-\frac{\partial V}{\partial z}+\sqrt{\eta k_{B}T}\zeta(t),
\end{equation}
where $\zeta(t)$ is a zero-mean Gaussian white noise with the correlation $\langle\zeta_{i}(t_{1})\zeta_{j}(t_{2})\rangle=2\delta_{ij}\delta(t_{1}-t_{2})$. The value of the slenderness parameter used in the numerical simulations is $\epsilon=0.01$.

\begin{figure}[!ht]
\centering
\includegraphics[width=4in]{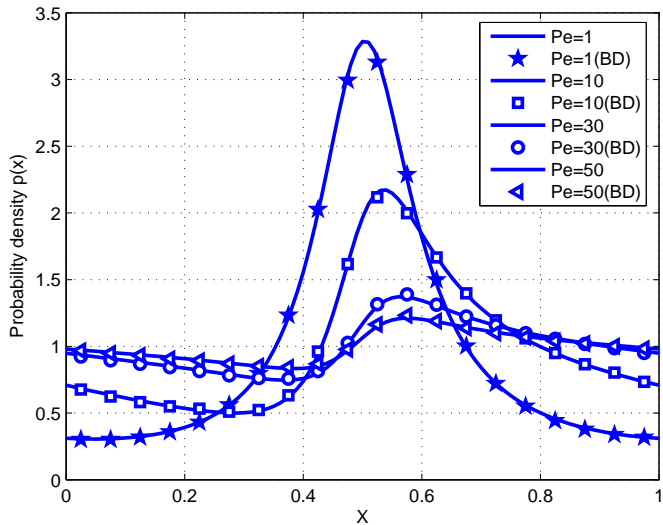}
\caption{Marginal probability density $p(x)$ for different values of the P\'eclet number ($\textrm{Pe}$). The solid lines indicate the asymptotic
results and the symbols correspond to the results of the Brownian dynamics simulations.}
\label{fig:p0x_both}
\end{figure}

\begin{figure}[!ht]
\centering
\includegraphics[width=4in]{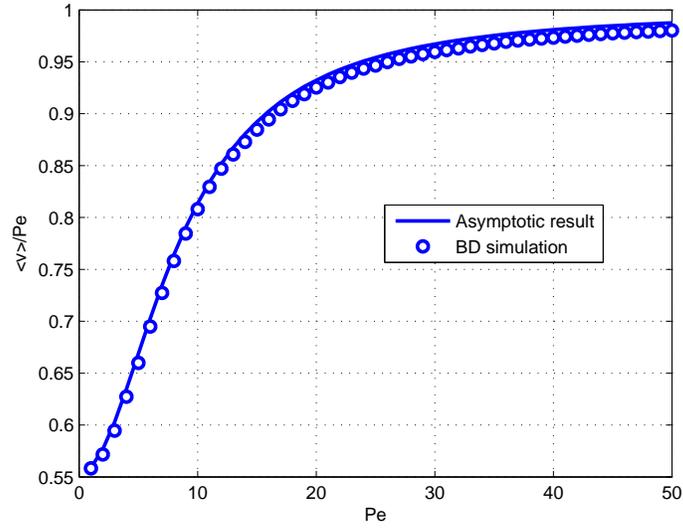}
\caption{Average velocity along the channel as a function of the P\'eclet number. The solid line corresponds to the leading order term of the
asymptotic results and the solid circles represent the values computed from the Brownian dynamics simulations.}
\label{fig:v_both}
\end{figure}

\begin{figure}[!ht]
\centering
\includegraphics[width=4in]{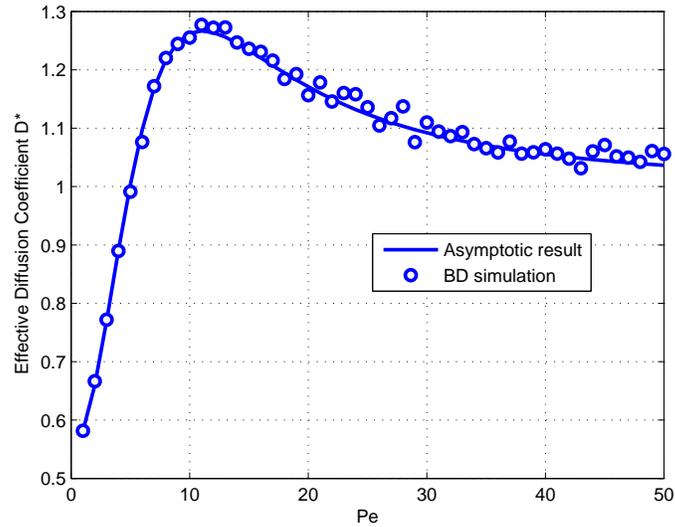}
\caption{Asymptotic dispersion coefficient as a function of the P\'eclet numbers. The solid line corresponds to the leading order term in the
asymptotic analysis. The solid circles corresponds to the dispersion coefficient calculated from the Brownian dynamics simulations.}
\label{fig:d_both}
\end{figure}

In figure \ref{fig:p0x_both} we show the marginal probability distribution $p(x)$ in a single period of the potential, for different values of the P\'eclet number $\textrm{Pe}=1, 10, 30, 50$. The leading order term of the distribution, given by Eq. (\ref{marginal}), agrees well with the results of the Brownian dynamics
simulations for all P\'eclet numbers. The figure also shows that the marginal probability distribution tends to a uniform distribution
as the P\'eclet number increases. This indicates that, as discussed before, the effect of the potential on the particle distribution, as well as the contribution
of diffusive transport to the average velocity of the particles, decreases as the P\'eclet number increases.
Figure \ref{fig:v_both} shows that the average velocity of the particles obtained in the simulations is accurately described by the
leading order term in the perturbation expansion. Finally, in figure \ref{fig:d_both} we compare the effective dispersion coefficient obtained from the leading
order term in the $B$-field (numerically solving Eq. (\ref{eqn:Bfield}) by means of finite differences) with that computed directly from the simulations.
We observe good agreement between the simulations and the asymptotic analysis.

\section{Conclusions}

We have investigated the transport of Brownian particles driven by a uniform external force. The particles are confined to a narrow periodic channel by a parabolic
potential in the cross-section. We used asymptotic methods to obtain the leading order solution of the two-dimensional Smoluchowski equation for the long-time
probability distribution of particles reduced to a single period of the potential. We first showed that the leading order analysis reproduces a previously proposed
extension of the Fick-Jacobs approximation to biased transport. We thus provide a systematic method to improve on the Fick-Jacobs approximation through higher order analysis. We also showed
that the leading order equation is equivalent to that obtained for solid channels and thus demonstrated that a confining potential has analogous effects on the
distribution of particles and their transport parameters, such as the average velocity and the asymptotic dispersion coefficient.
We then analyzed the case of a cosine variation in the aperture of the confining channel and compared the results with Brownian dynamics simulations.
We compared the long-time marginal distribution as well as its first moments (average velocity and dispersion coefficient) and obtained excellent agreement with the first order in the asymptotic analysis over a wide range of P\'eclet numbers.

This material is partially based upon work supported by the National Science Foundation under
Grant No. CBET-0731032.

\end{document}